\documentstyle[11pt,epsfig]{article}
 
\def\ga{\gamma}
\def\als{{\alpha_s}}

\newcommand{\gapproxeq}{\lower .7ex\hbox{$\;\stackrel{\textstyle >}{\sim}\;$}}
\newcommand{\lapproxeq}{\lower .7ex\hbox{$\;\stackrel{\textstyle <}{\sim}\;$}}
\newcommand{\bee}{\begin{equation}}
\newcommand{\ene}{\end{equation}}
\newcommand{\bea}{\begin{eqnarray}}
\newcommand{\ena}{\end{eqnarray}}
\newcommand{\bear}  {\begin{array}}
\newcommand{\enar}  {\end{array}}

\newcommand{\uupv}{u^{\uparrow}_{val}}
\newcommand{\udwv}{u^{\downarrow}_{val}}
\newcommand{\dupv}{d^{\uparrow}_{val}}
\newcommand{\ddwv}{d^{\downarrow}_{val}}

\newcommand{\fnx}{{F_{2}^{n}}(x)}
\newcommand{\fpx}{{F_{2}^{p}}(x)}
\newcommand{\gpx}{{g_{1}^{p}}(x)}

\newcommand{\gdx}{{g_{1}^{d}}(x)}

\begin{document}

{{\hfill \small $\begin{array}{r}\mbox{
Universit\'a di Napoli Preprint DSF-33/96 (july 1996)} \\
\mbox{hep-ph/9607338}\end{array}$}}

\begin{center}
{\Large\bf Statistical Inspired Parton Distributions and the Violation of QPM
Sum Rules}
\end{center}
\bigskip\bigskip

\begin{center}
{{\bf F. Buccella}$^{\dagger}$, {\bf I. Dor\u{s}ner}$^{\dagger \ddagger}$, {\bf
G. Miele}$^{\dagger}$, {\bf O. Pisanti}$^{\dagger}$, {\bf P.
Santorelli}$^{\dagger}$, and {\bf N. Tancredi}$^{\dagger}$} 
\end{center}

\bigskip

\begin{itemize}
\item[$\dagger$] {\it Dipartimento di Scienze Fisiche, Universit\`a di Napoli
''Federico II'', and INFN Sezione di Napoli, Mostra D'Oltremare Pad. 20,
I--80125 Napoli, Italy} 
\item[$\ddagger$] {\it Prirodno-Matemati\u{c}ki Fakultet, Univerzitet u
Sarajevu, Zmaja od Bosne 32, Sarajevu, R Bosna i Hercegovina} 
\end{itemize}

\bigskip\bigskip\bigskip

\begin{abstract}
A quantum statistical parametrization of parton distributions has been
considered. In this framework, the exclusion Pauli principle connects the
violation of the Gottfried sum rule with the Ellis and Jaffe one, and
implies a defect in the Bjorken sum rule. However, in terms of standard
parametrizations of the polarized distributions a good description of the
data is obtained once a large gluon polarization is provided.
Interestingly, in this description there is no violation of the Bjorken sum
rule. 
\end{abstract}

\baselineskip22pt

\section{Statistical distributions for partons}

The experimental results on deep inelastic scattering (DIS) are always an
inexhaustible source for deeper insight in the nucleon structure. Among
them, the violation of well established sum rules represents the relevant
starting point to unveil the mechanisms which rule the parton physics. In
particular, the value found by the EMC experiment \cite{emc} for the first
moment of the structure function $g_1^p (x)$, $\Gamma_1^p = 0.126\pm
0.010\pm 0.015$, results to be smaller than the prediction based on the
Ellis and Jaffe sum rule (EJ), $I_p^{EJ} = (F/2) - (D/18)$. For this reason
it has generated the so-called {\it spin crisis}.\\ One of the possible
explanations of the EMC measurement is that it is a consequence of a
negative gluon contribution \cite{anom}, $-(\als/6 \pi) \Delta G$, with the
large positive value of $\Delta G$ balanced by a negative and large $L_z$.
The gluonic contribution, which is related to the axial anomaly, is the
same for proton and neutron and would not affect the Bjorken sum rule (Bj).

An alternative interpretation of the defect in the EJ for the proton can be
found if Pauli principle plays an important role in parton distributions.
Remarkably, these considerations lead to connect \cite{bs} the violation of
the EJ with the observed defect in the Gottfried sum rule. 

According to Feynman and Field \cite{Fey}, in the proton $\bar u < \bar d~$
 since only one valence $d$ quark is present with respect to the two $u$
quarks. This hypothesis would be confirmed both from the NMC measurements
\cite{f2p-3g-2} of the Gottfried sum rule and from the CERN-NA51 experiment
\cite{NA51} on dilepton Drell-Yan production in $p p$ and $p n$ reactions.
As far as the Gottfried sum rule is concerned, the NMC collaboration
measures 
\bee
\int_0^1 {dx \over x} \left(\fpx - \fnx\right) = 0.235\pm 0.026,
\ene
which, together with the Adler sum rule, implies $\bar u - \bar d = -
0.15\pm 0.04$ and $u - d = 0.85 \pm 0.04$. 

A further confirmation of the Feynman and Field conjecture comes from the
experimental observation that, at high $x$, $\fnx/\fpx \rightarrow 1/4$ and
$A_1^p \rightarrow 1$. This feature, in fact, suggests that $\uupv$ is the
dominating parton distribution at high $x$. Indeed, at $Q^2 =0$, the axial
couplings of the baryon octet are fairly described in terms of the valence
quarks 
\bee
\uupv  = 1+F~~~,~~~\udwv  = 1-F~~~,~~~\dupv  = {1+F-D \over 2}~~~,
~~~\ddwv  = {1-F+D \over 2}~~~. 
\ene
Thus, by using the experimental values for $F$ and $D$ \cite{FD}, one 
gets
\bee
\uupv \simeq { 3 \over 2} \simeq \udwv + \dupv + \ddwv~~~.
\ene
Hence, the correlation {\it broader shapes $\leftrightarrow$ higher
abundances}, just dictated by Pauli principle, is very well verified by the
experimental results. For these reasons, it is natural to assume
Fermi--Dirac distributions in the variable $x$ for the quark partons 
\bee
p(x) = f(x) \left[\exp\left({ x - \tilde{x}(p) \over \bar{x}} \right)
+ 1 \right]^{-1}~~~, 
\ene
where $\bar{x}$ plays the role of the temperature, $\tilde{x}(p)$ is the
{\it thermodynamical potential} of the parton $p$, identified  by its
flavour and spin direction, and $f(x)= A~ x^{\alpha} (1 -x)^{\beta}$ is a
weight function. Analogously, for the gluons one has 
\bee
G^{\uparrow(\downarrow)}(x) = {8 \over 3} f(x) 
\left[\exp\left({ x - \tilde{x}(G^{\uparrow(\downarrow)}) \over \bar{x}} 
\right) - 1 \right]^{-1}~~~.
\ene
To recover the power behaviour of $\fpx$ at small $x$ we add a {\it liquid}
unpolarized component for the light quark--partons ($u$, $d$ and their
antiparticles), $f_L(x) = (A_{L}/2)~x^{\alpha_{L}} (1-x)^{\beta_L}$, and
the same $x$ dependence, but with a different normalization for $s$ and
$\bar{s}$. With these distributions we try to reproduce \cite{bmt} the
structure functions $\fpx$, $\fnx$ \cite{f2p-3g-2,f2nf2p}, $xF_3(x)$
\cite{xf3-3g}, $\gpx$ and $\gdx$ \cite{g1slac,g1cern}, at $Q^2 = 3$ and
$10~(GeV/c)^2$, considering the options with polarized or unpolarized
gluons. We report in Tables I.a and I.b the parameters and the gas
abundances for partons, found with $\Delta \bar{q}_i=0$ and with/without
$\Delta G(x)$ at $Q^2 = 3$, and $10~(GeV/c)^2$, and compare them with the
results of a previous analysis. Figures 1.a--5.a and 1.b--5.b show our
theoretical predictions versus the experimental data corresponding to
$Q^2=3$ and $Q^2=10~(GeV/c)^2$ respectively, for the fits with $\Delta G=0$
(solid line) and $\Delta G \neq 0$ (dashed line) reported in Tables I.a and
I.b. 

\section{Standard parametrizations for the polarized distributions}

In order to test in a model independent way the above two interpretations
for the violation of polarized sum rules, we consider \cite{bbpss} a
standard parametrization \cite{gerst}, at $Q_0^2= 3\,(GeV/c)^2$, 
\bee
\bear{lcl}
x \Delta u_v (x, Q_0^2) &=& \eta_u A_u x^{a_u} (1-x)^{b_u} (1+\ga_u x),
\vspace{.2truecm} \\
x \Delta d_v (x, Q_0^2) &=& \eta_d A_d x^{a_d} (1-x)^{b_d} (1+\ga_d x),
\vspace{.2truecm} \\
x \Delta G (x, Q_0^2)   &=& \eta_G A_G x^{a_G} (1-x)^{b_G} (1+\ga_G x),
\enar
\ene
with $A_q = A_q(a_q,b_q,\ga_q$) ($q~=~u,d,G$), in such a way that
\bee
\int_0^1 dx A_q x^{a_q-1} (1-x)^{b_q} (1+\ga_q x) = 1.
\ene

We fix $\eta_d(Q_0^2) = \tilde F - \tilde D = -0.24\pm 0.04$ and explore
the two options {\bf A} and {\bf B}, the first one with $\eta_u(Q_0^2) =
2~\tilde F = 0.78\pm 0.03$ and $\eta_G$ free, the second one with $\eta_u$
free and $\eta_G = 0$. Options {\bf A} and {\bf B} correspond to the
interpretation of the defect in the EJ for proton in terms of the anomaly,
assuming that the Bj is obeyed, and to the case of a smaller $\Delta u$
resulting from the Pauli principle, respectively. 

The parameters corresponding to the best fit of the SLAC proton and
deuteron data \cite{g1slac} for the options {\bf A} and {\bf B} are given
in Table II, while in Figs. 6.a and 6.b one compares the two resulting
curves with SLAC data at $<\!\!Q^2\!\!> = 3\,(GeV/c)^2$ with proton and
deuteron targets; for the later case we take 
\bee
g_1^d(x) = \frac{1}{2} \left ( 1 - \frac{3}{2} \omega_D \right) 
(g_1^p(x) + g_1^n(x)),
\ene
to account for the small D-wave component in the deuteron ground state,
with $\omega_D = 0.058$. The evolution of the results for the two options
to 2 and $10~(GeV/c)^2$, obtained by numerically solving the
Altarelli-Parisi equations, is compared with the SLAC \cite{E142} and CERN
data \cite{g1cern} respectively in Figs. 7.a, 7.b, and 7.c. 

\section{Conclusions}

Quantum statistical inspired distributions for quarks and gluons provide a
fair description of the experimental data on deep inelastic scattering in
terms of few free parameters. In this approach, the parton distributions
are given in terms of a universal {\it weight} function, which accounts for
the parton density levels at fixed $x$, {\it thermodynamical potentials}
and of a quantity which plays the role of {\it temperature}. Furthermore,
the violations of the Gottfried and EJ sum rules result to be connected and
both imply a defect in the Bj. 

As far as a standard parametrization of the polarized distributions is
concerned, it is worth stressing the value $\eta_u = 0.63 \pm 0.03$ for
option {\bf B}, smaller than $2 \tilde F = 0.78\pm 0.03$ from quark parton
model, as predicted by Pauli principle. It is interesting to remark that
with both options one fails to reproduce the rise of $x g_1^p (x)$ at small
$x$, which is certainly welcome to increase the contribution to the l.h.s.
of the Bj.

%\bigskip\bigskip
\par\noindent
\begin{center}
{\bf Table I.a}\\
\bigskip
{\small
\begin{tabular}{|c|c|c|c|c|c|c|}
\hline
Param.
& \multicolumn{2}{c|}{Previous fit} & 
\multicolumn{2}{c|}{Present fit with only} &
\multicolumn{2}{c|}{Present fit with only} \\
$Q^2=3$& \multicolumn{2}{c|}{$~$} & 
\multicolumn{2}{c|}{$\Delta u$, $\Delta d \neq 0$} &
\multicolumn{2}{c|}{$\Delta u$, $\Delta d$, $\Delta G \neq 0$}\\
$(GeV/c)^2$& \multicolumn{2}{c|}{($\chi_{NDF}^2=2.47$)} & 
\multicolumn{2}{c|}{($\chi_{NDF}^2=2.33$)} &
\multicolumn{2}{c|}{($\chi_{NDF}^2=2.32$)}\\
\hline
$A$ & \multicolumn{2}{c|}{$ 2.66\begin{array}{c} +.09\\ -.08 \end{array} $}
& \multicolumn{2}{c|}{$ 2.51\pm .07$}
& \multicolumn{2}{c|}{$ 2.54\pm .08$}\\
$\alpha$ & \multicolumn{2}{c|}{$ -.203 \pm .013$}
& \multicolumn{2}{c|}{$ -.231 \pm .012$}
& \multicolumn{2}{c|}{$ -.231 \pm .011$}\\
$\beta$ &\multicolumn{2}{c|}{$ 2.34 \begin{array}{c} +.05\\ -.06\end{array}$}
&\multicolumn{2}{c|}{$ 2.21 \pm .04$}
&\multicolumn{2}{c|}{$ 2.22 \pm .04$} \\
$A_{L}$ 
& \multicolumn{2}{c|}{$ .0895 \begin{array}{c} +.0107\\-.0084\end{array}$}
& \multicolumn{2}{c|}{$ .127 \begin{array}{c} +.016\\-.013\end{array}$}
& \multicolumn{2}{c|}{$ .128 \begin{array}{c} +.015\\-.013\end{array}$}\\
$\alpha_{L}$ & \multicolumn{2}{c|}{$ -1.19 \pm .02$}
& \multicolumn{2}{c|}{$ -1.18 \begin{array}{c} +.03\\-.02\end{array}$}
& \multicolumn{2}{c|}{$ -1.18 \pm .02$}\\
$\beta_{L}$&\multicolumn{2}{c|}{$7.66 \begin{array}{c}+1.82\\-1.59
\end{array}$}
&\multicolumn{2}{c|}{$10.3 \begin{array}{c}+1.4\\-1.3\end{array}$}    
&\multicolumn{2}{c|}{$10.1 \begin{array}{c}+1.4\\-1.3\end{array}$}\\    
\cline{2-7}
$\bar{x}$&$.235 \pm .009$ & gas abund. &$.214 \pm .008$ & gas abund.
&$.223 \pm .011$ & gas abund. \\
\cline{3-3} \cline{5-5}  \cline{7-7}
$\tilde{x}(u^{\uparrow})$& $1.00 \pm .07$ & $1.15 \pm .01$
& $1.00 \pm .02$ & $1.22 \pm .01$
& $1.00 \pm .02$ & $1.23 \pm .01$\\
$\tilde{x}(u^{\downarrow})$ & $.123 \pm .012$ & $.53 \pm .01$
& $.141 \pm .011$ & $.575 \pm .009$
& $.129 \begin{array}{c}+.014\\-.015\end{array}$ & $.566 \pm .019$\\
$\tilde{x}(d^{\uparrow})$& $-.068 \begin{array}{c} +.021\\
-.024\end{array}$ & $.33 \pm .03$ & $-.029 \begin{array}{c} +.019\\
-.020\end{array}$ & $.366 \pm .025$ & $-.028 \pm .020$ & $.379 \pm .030$ \\
$\tilde{x}(d^{\downarrow})$ & $.200 \begin{array}{c} +.013\\
-.014\end{array}$ & $.62 \pm .01$ & $.211 \pm .011$ & 
$.667 \pm .006$ & $.196 \begin{array}{c} +.015\\
-.016\end{array}$ & $.651 \pm .017$\\
$\tilde{x}(\bar{u}^{\uparrow})$
& $-.886 \pm .266$ & $.015 \begin{array}{c} +.034\\ -.009\end{array}$
& $-.522 \begin{array}{c}+.049\\-.061\end{array}$ 
& $.054 \begin{array}{c} +.021\\ -.022\end{array}$    
& $-.559 \begin{array}{c}+.057\\-.075\end{array}$ 
& $.052 \pm .022$ \\
$\tilde{x}(\bar{u}^{\downarrow})$& $''$ & $''$ & $''$ & $''$
& $''$ & $''$\\
$\tilde{x}(\bar{d}^{\uparrow})$
& $-.460\begin{array}{c} +.047\\
-.064\end{array}$ &$.08 \begin{array}{c} +.03\\-.02\end{array}$
& $-.339\begin{array}{c} +.032\\
-.040\end{array}$ &$.12 \pm .03$
& $-.366\begin{array}{c} +.037\\
-.049\end{array}$ &$.12 \pm .03$\\
$\tilde{x}(\bar{d}^{\downarrow})$ & $''$ & $''$ & $''$ & $''$ & $''$ & $''$\\
$\tilde{x}(G^{\uparrow})$
& $-.067$ & $3.16$ & $-.067\begin{array}{c} +.008 \\ -.009\end{array}$ 
& $2.93 \pm .40$
& $-.069\pm .09$ 
& $3.04 \pm .55$\\
$\tilde{x}(G^{\downarrow})$ & $''$ & $''$ & $''$ & $''$ 
& $-.085\begin{array}{c} +.015 \\ -.019\end{array}$  
& $2.56 \pm .61$\\
\hline
\end{tabular}}
\end{center}
\footnotesize{The parameters and the gas abundances for partons, 
found with $\Delta \bar{q}_i=0$ and with/without $\Delta G(x)$ at 
$Q^2 = 3~(GeV/c)^2$ are reported and compared with the results of 
a previous analysis. Note that, no antiquarks or strange quark polarization
is assumed.}
\newpage
\bigskip\bigskip
\par\noindent
\begin{center}
{\bf Table I.b}\\
\bigskip
{\small 
\begin{tabular}{|c|c|c|c|c|}
\hline
Parameters & \multicolumn{2}{c|}{Present fit with only} &
\multicolumn{2}{c|}{Present fit with only} \\
$Q^2=10$& \multicolumn{2}{c|}{$\Delta u$, $\Delta d \neq 0$} &
\multicolumn{2}{c|}{$\Delta u$, $\Delta d$, $\Delta G \neq 0$}\\
$(GeV/c)^2$& \multicolumn{2}{c|}{($\chi_{NDF}^2=0.98$)} &
\multicolumn{2}{c|}{($\chi_{NDF}^2=0.95$)}\\
\hline
$A$ & \multicolumn{2}{c|}{$ 2.00\begin{array}{c} +.15\\ -.13 \end{array} $}
& \multicolumn{2}{c|}{$ 1.99\begin{array}{c} +.19\\ -.14 \end{array} $}\\
$\alpha$ 
& \multicolumn{2}{c|}{$ -.363 \begin{array}{c} +.042\\ -.035 \end{array} $}
& \multicolumn{2}{c|}{$ -.375 \begin{array}{c} +.046\\ -.035 \end{array}$}\\
$\beta$ 
&\multicolumn{2}{c|}{$ 2.29 \pm .04$}
&\multicolumn{2}{c|}{$ 2.28 \begin{array}{c} +.05 \\-.04 \end{array}$} \\
$A_{L}$ 
& \multicolumn{2}{c|}{$ .108 \begin{array}{c} +.018\\-.014\end{array}$}
& \multicolumn{2}{c|}{$ .109 \begin{array}{c} +.022\\-.016\end{array}$}\\
$\alpha_{L}$ 
& \multicolumn{2}{c|}{$ -1.29 \pm .02$}
& \multicolumn{2}{c|}{$ -1.28 \begin{array}{c} +.03\\ -.02 \end{array}$}\\
$\beta_{L}$
&\multicolumn{2}{c|}{$10.1 \begin{array}{c}+3.7\\-3.5\end{array}$}    
&\multicolumn{2}{c|}{$11.1 \begin{array}{c}+4.7\\-4.3\end{array}$}\\    
\cline{2-5}
$\bar{x}$ &$.238 \begin{array}{c} +.009\\ -.007 \end{array} $ & gas abund.
&$.241 \begin{array}{c} +.030\\ -.020 \end{array}$ & gas abund. \\
\cline{3-3} \cline{5-5}  
$\tilde{x}(u^{\uparrow})$& $1.00 \pm .01$ & $1.32 \begin{array}{c} +.05\\ -.06 
\end{array}$
& $1.00 \pm .01$ & $1.35 \begin{array}{c} +.05\\ -.07 \end{array}$\\
$\tilde{x}(u^{\downarrow})$ 
& $.104 \begin{array}{c} +.038\\ -.043 \end{array} $ 
& $.617 \begin{array}{c} +.012\\ -.030 \end{array}$ & $.092 
\begin{array}{c} +.046\\ -.071 \end{array}$ 
& $.62 \begin{array}{c} +.03\\ -.08 \end{array}$\\
$\tilde{x}(d^{\uparrow})$ & $-.114 \begin{array}{c} +.050\\
-.070\end{array}$ & $.36 \begin{array}{c} +.04\\ -.06 \end{array}$ 
& $-.082 \begin{array}{c} +.056\\
-.091\end{array}$ & $.41 \begin{array}{c} +.06\\ -.10 \end{array}$ \\
$\tilde{x}(d^{\downarrow})$ & $.171 \begin{array}{c} +.022\\
-.027\end{array}$ & $.704 \begin{array}{c} +.014\\ -.003 \end{array}$ 
& $.143 \begin{array}{c} +.034\\
-.054\end{array}$ & $.69 \begin{array}{c} +.03\\ -.05 \end{array}$\\
$\tilde{x}(\bar{u}^{\uparrow})$
& $-.69 \begin{array}{c} +.13\\ -.20 \end{array}$ & $.044 
\begin{array}{c} +.067\\ -.049\end{array}$    
& $-.65 \begin{array}{c} +.15\\ -.26 \end{array}$ & $.055 
\begin{array}{c} +.064\\-.041\end{array} $ \\
$\tilde{x}(\bar{u}^{\downarrow})$& $''$ & $''$ & $''$ & $''$\\
$\tilde{x}(\bar{d}^{\uparrow})$ & $-.412\pm.12$ 
& $.130 \begin{array}{c} +.001\\ -.025 \end{array}$
& $-.40\pm.10$ &$.14 \begin{array}{c} +.08\\ -.06 \end{array}$\\
$\tilde{x}(\bar{d}^{\downarrow})$& $''$ & $''$ & $''$ & $''$\\
$\tilde{x}(G^{\uparrow})$
& $-.086\pm 0.003$ &$3.519 \begin{array}{c} +.001\\ -.025 \end{array}$
& $-.076 \begin{array}{c} +.013\\-.020\end{array}$ 
&$4.2 \pm 1.0$\\
$\tilde{x}(G^{\downarrow})$& $''$ & $''$ 
& $-.108\begin{array}{c} +.027 \\ -.056\end{array}$  & 
$3.0 \pm 1.2$\\
\hline
\end{tabular}}
\end{center}
{\footnotesize{The same quantities of Table I.a are shown for 
$Q^2 = 10~(GeV/c)^2$.}}

\begin{center}
{\bf Table II}\\
\bigskip
\begin{small}
\begin{tabular}{|c|c|c|c|c|c|c|c|} \hline\hline
\rule[-0.4truecm]{0mm}{1truecm}
& $a_u=a_d$ & $b_u$ & $\gamma_u=\gamma_d$ & $b_d$ & $a_G$ & $b_G$ & $\gamma_G$ 
\\
\hline\hline
\rule[-0.4truecm]{0mm}{1truecm}
{\bf A} & $0.52\pm 0.09^{(*)}$ & $1.7\pm 0.3^{(*)}$ & $2.8_{-1.9}^{+3.7(*)}$ & 
$3.0\pm 0.5^{(*)}$ & 1 & $20\pm 11^{(*)}$ & $0\pm 14^{(*)}$ \\
\hline
\rule[-0.4truecm]{0mm}{1truecm}
& $\eta_u=0.78\pm 0.03$ & $\eta_d=-0.24\pm 0.02$ & $\eta_G=1.1\pm 0.5^{(*)}$ 
& $\chi_{NDF}^2=1.04$ &&& \\
\hline\hline
\rule[-0.4truecm]{0mm}{1truecm}
{\bf B} & $1.0\pm 0.1^{(*)}$ & $1.8_{-0.2}^{+0.5(*)}$ & $0.0\pm 2.7^{(*)}$ & 
$4.1\pm 1.1^{(*)}$ & - & - & - \\
\hline
\rule[-0.4truecm]{0mm}{1truecm}
& $\eta_u=0.63\pm 0.03^{(*)}$ & $\eta_d=-0.24\pm 0.02$ & $\eta_G=0$ 
& $\chi_{NDF}^2=1.08$ &&& \\
\hline
\end{tabular}
\end{small}
\end{center}
{\footnotesize{The results of the options {\bf A} and {\bf B} (see text) for
the values of the parameters of the fits at $Q^2 = 3\,(GeV/c)^2$. The free
parameters are marked with an asterisk.}} 

\newpage
%-----------------------------------------------------------------------
\begin{figure}[t]
\epsfig{file=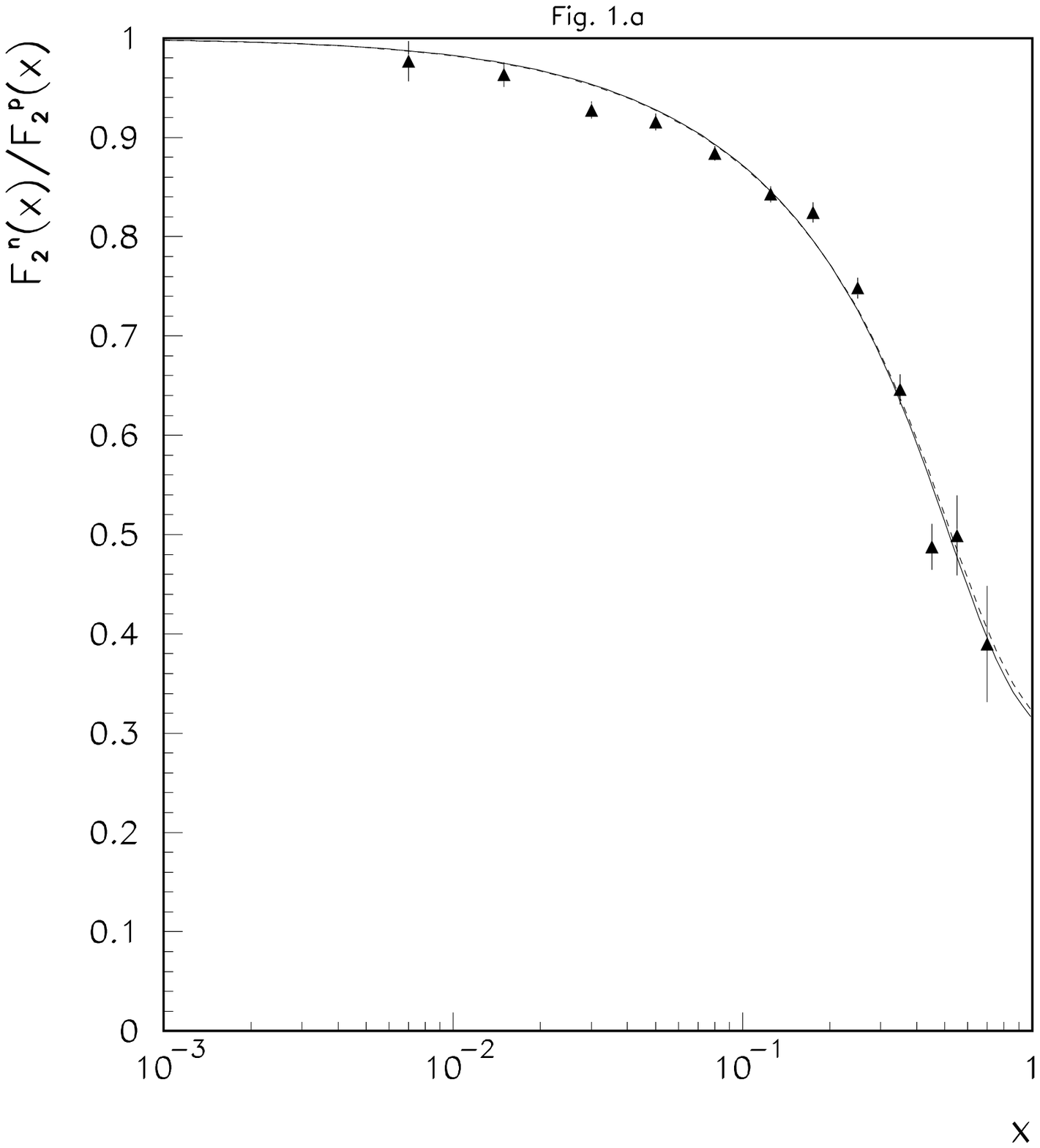,height=9cm}\quad
\epsfig{file=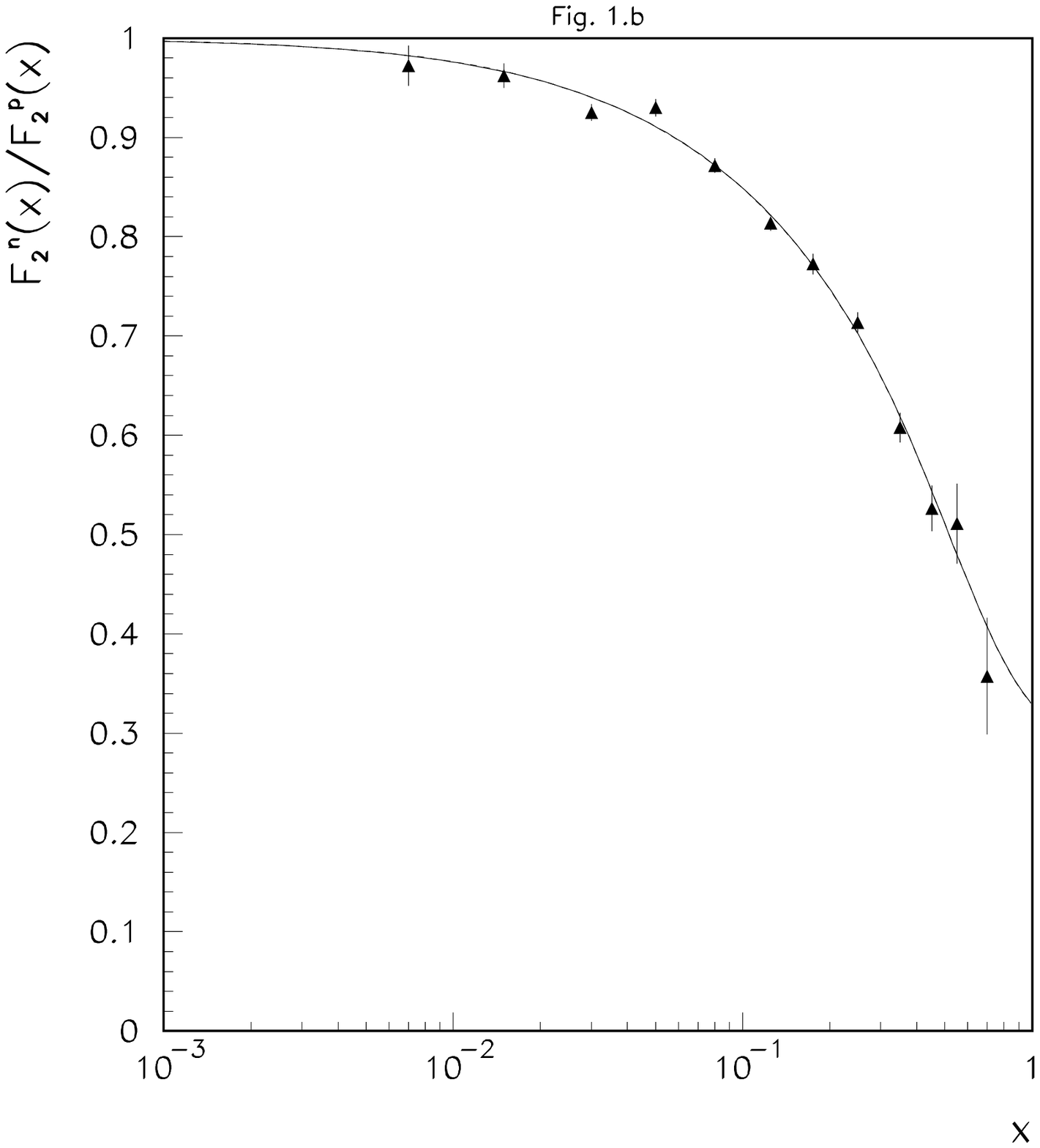,height=9cm}\\
\footnotesize{ 
\begin{itemize}
\item[Figure 1.a] The prediction for $\fnx/\fpx$ at $Q^2=3~(GeV/c)^2$ is
plotted and compared with the experimental data \cite{f2nf2p}, the solid
line and the dashed line corresponds to the fit with $\Delta G=0$ and
$\Delta G\neq0$ of Table I.a, respectively. This notation is valid for all
Figures 1.a-5.a. 
\item[Figure 1.b] The same quantity of Figure 1.a is plotted for
$Q^2=10~(GeV/c)^2$, the solid line and the dashed line corresponds to the
fit with $\Delta G=0$ and $\Delta G\neq0$ of Table I.b, respectively. This
notation is valid for all Figures 1.b-5.b. 
\end{itemize}}

\epsfig{file=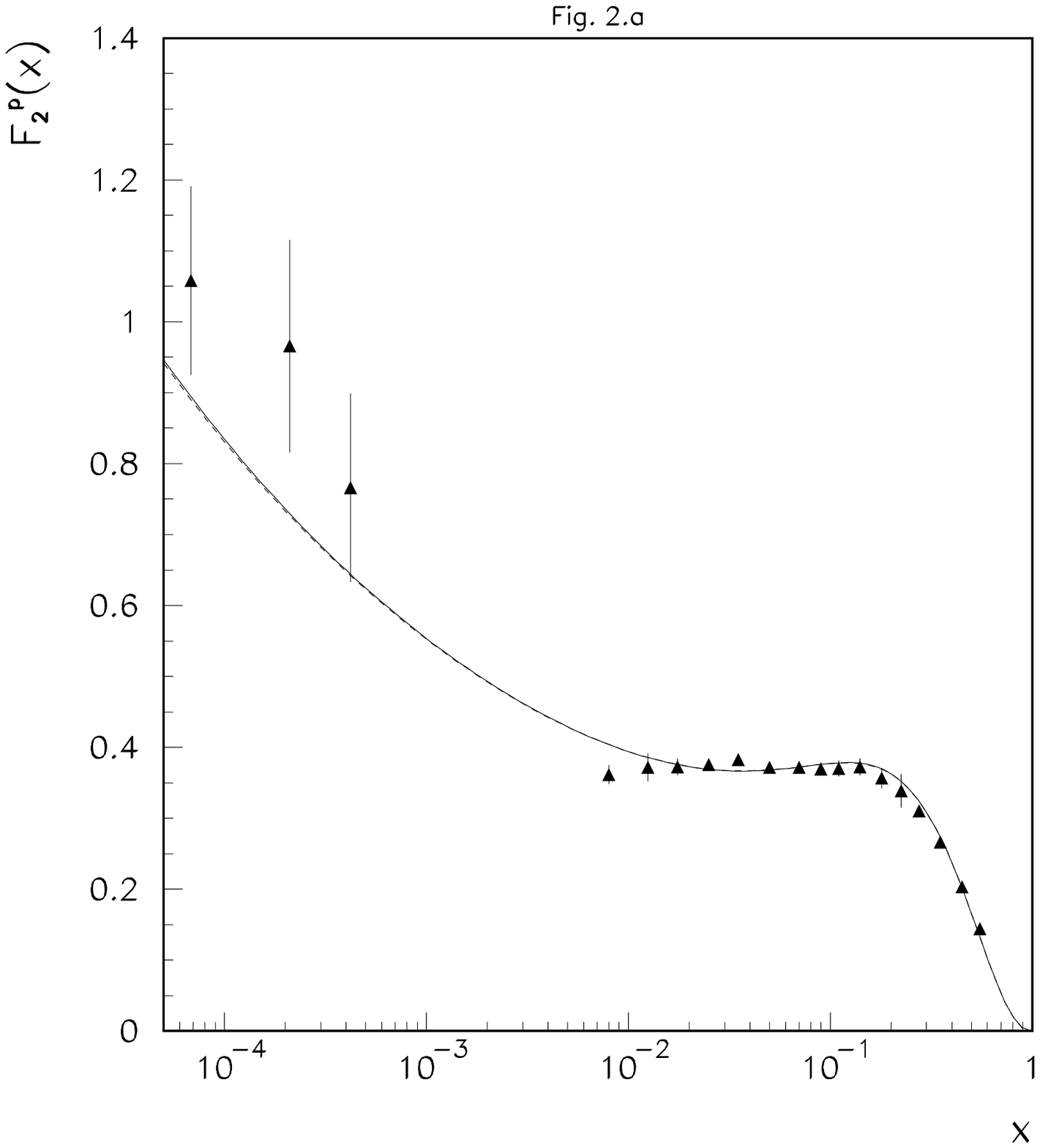,height=9cm}
\epsfig{file=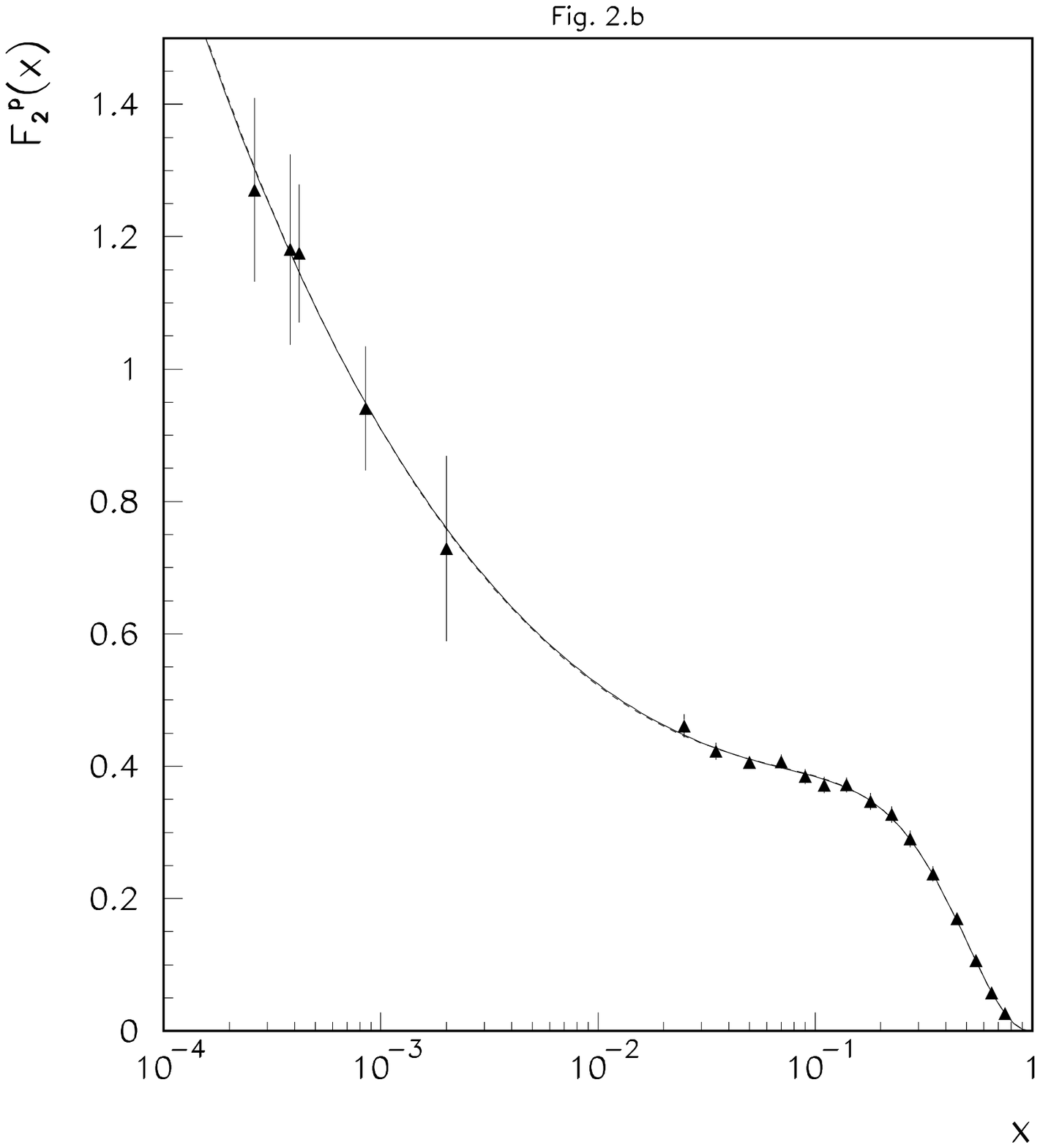,height=9cm}\\
\footnotesize{ 
\begin{itemize}
\item[Figure 2.a] The prediction for $\fpx$ at $Q^2=3~(GeV/c)^2$ is plotted
and compared with the experimental data \cite{f2nf2p}. 
\item[Figure 2.b] The same quantity of Figure 2.a is plotted for
$Q^2=10~(GeV/c)^2$. 
\end{itemize}}
\label{fig:fig1}
\end{figure}
%-----------------------------------------------------------------------
\newpage
\begin{figure}[t]
\epsfig{file=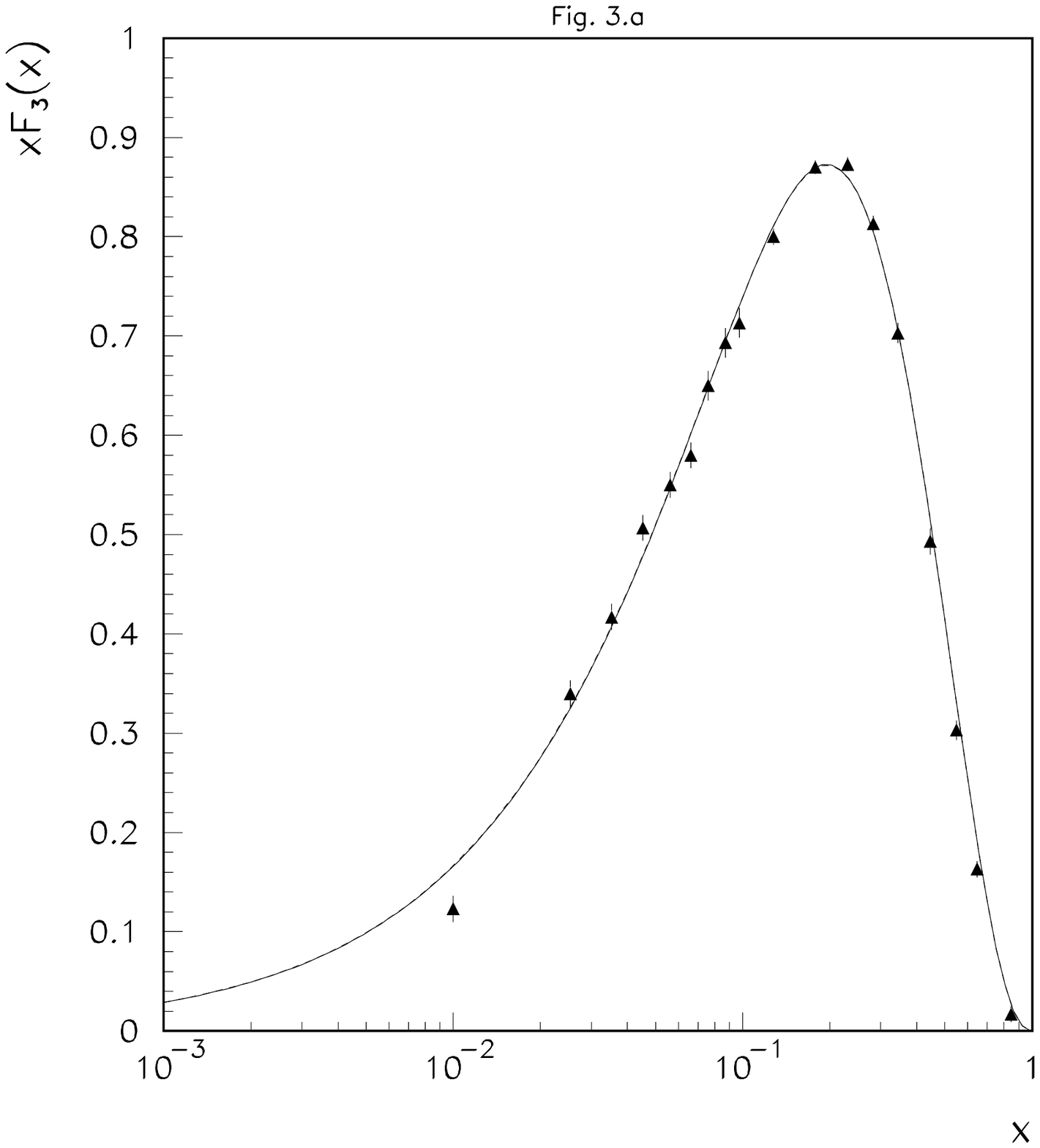,height=9cm}
\epsfig{file=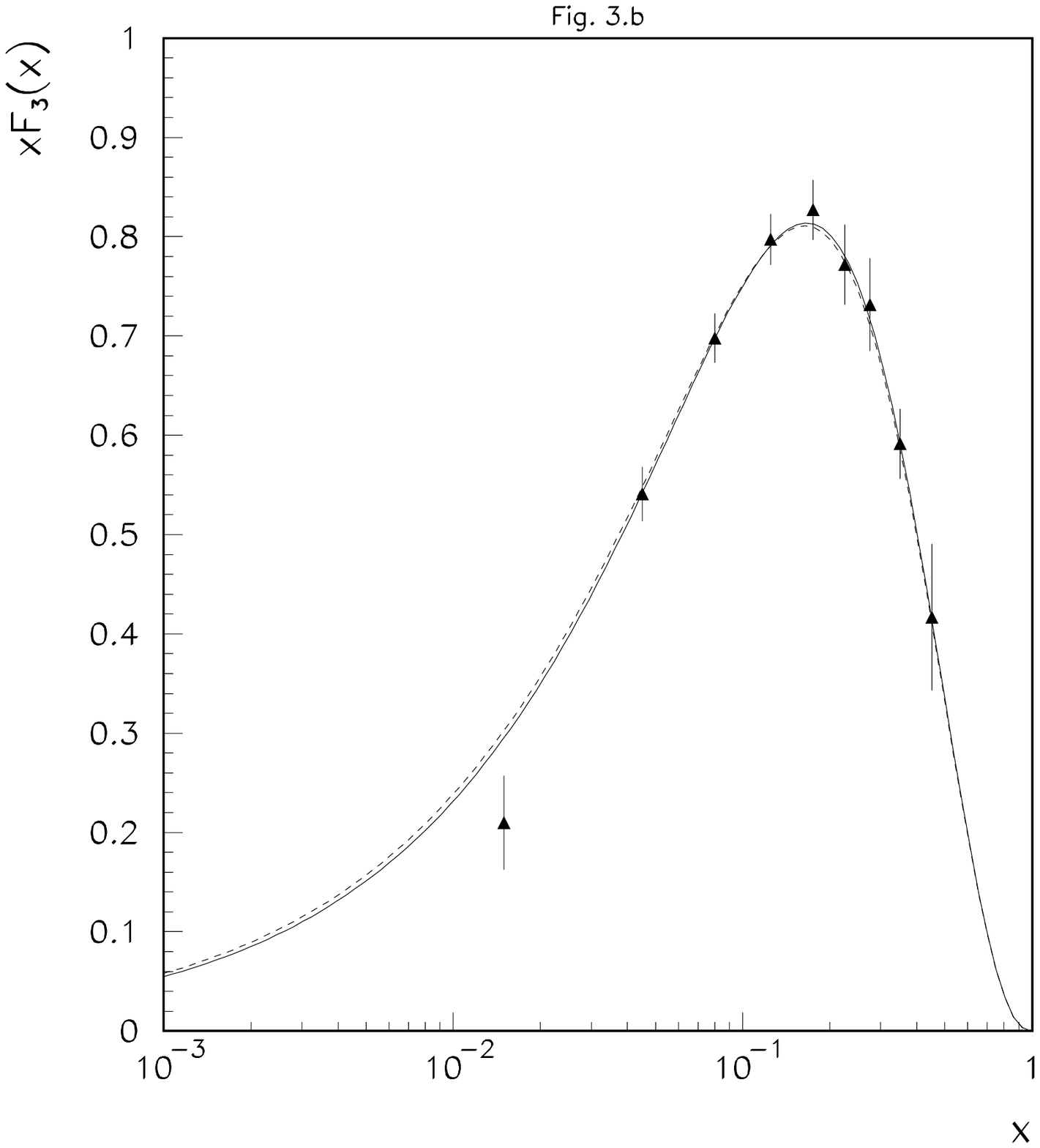,height=9cm}\\
\footnotesize{\begin{itemize}
\item[Figure 3.a] $xF_{3}(x)$ is plotted for $Q^2=3~(GeV/c)^2$ and the
experimental values are taken from \cite{xf3-3g}. 
\item[Figure 3.b] $xF_{3}(x)$ is plotted for $Q^2=10~(GeV/c)^2$ and the
experimental values are taken from \cite{xf3-3g}. 
\end{itemize}}

\epsfig{file=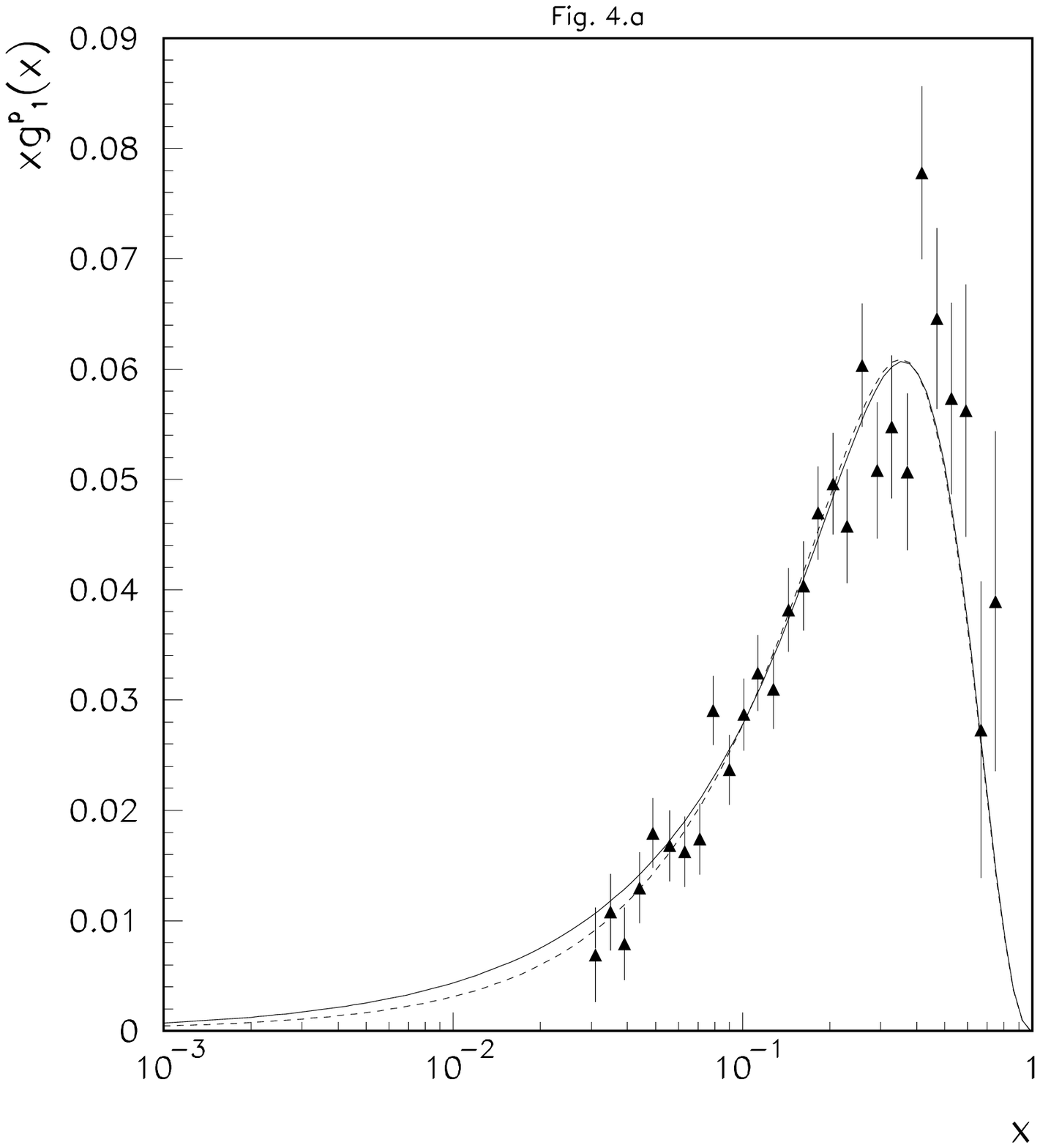,height=9cm}
\epsfig{file=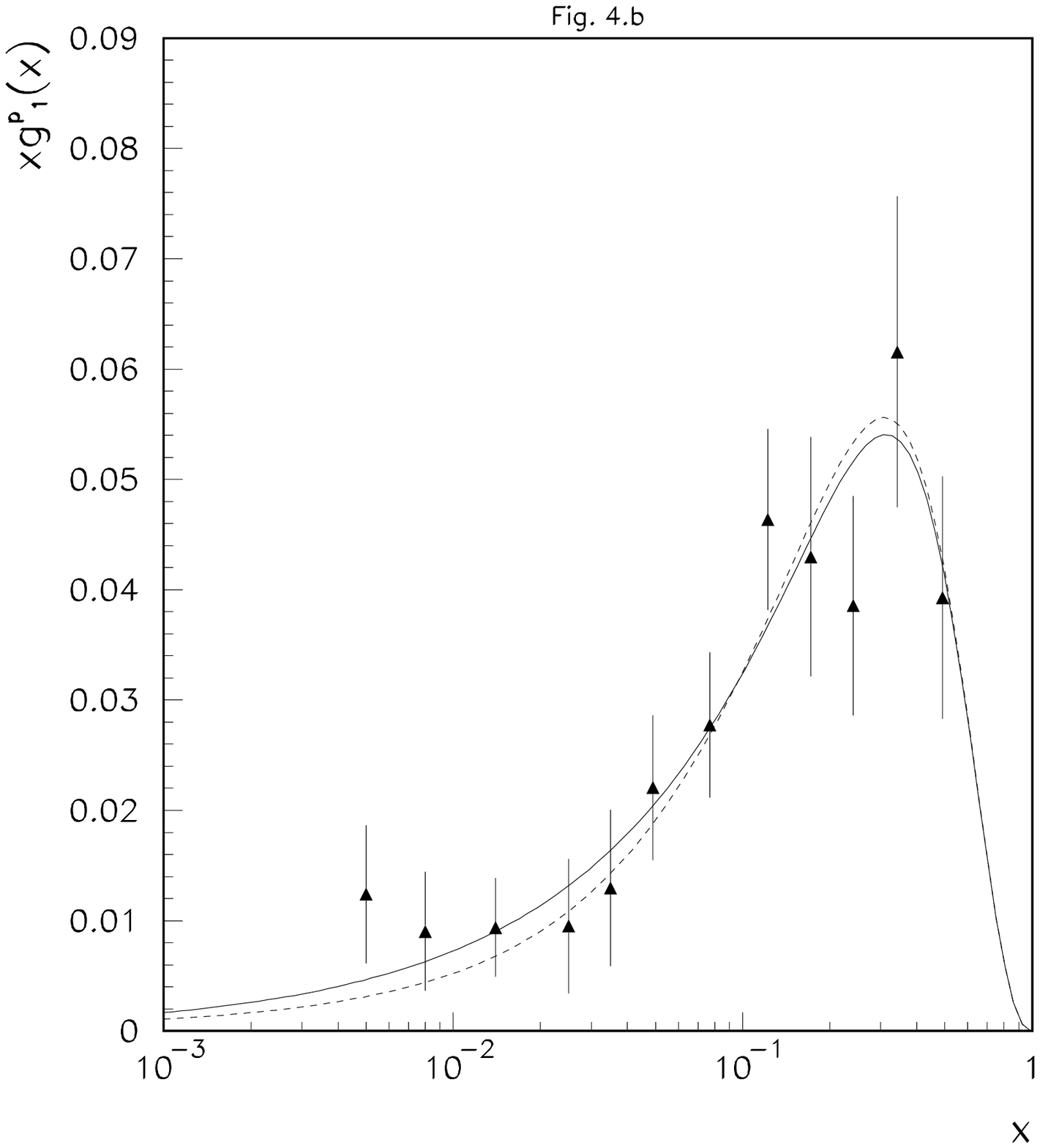,height=9cm}\\
\footnotesize{\begin{itemize}
\item[Figure 4.a] $x\gpx$ at $Q^2=3~(GeV/c)^2$ is plotted and compared with
the data \cite{g1slac}. 
\item[Figure 4.b] The same quantity of Figure 4.a corresponding to
$Q^2=10~(GeV/c)^2$ is plotted versus the experimental data \cite{g1cern}. 
\end{itemize}}
\label{fig:fig2}
\end{figure}
%-----------------------------------------------------------------------
\newpage
\begin{figure}[t]
\epsfig{file=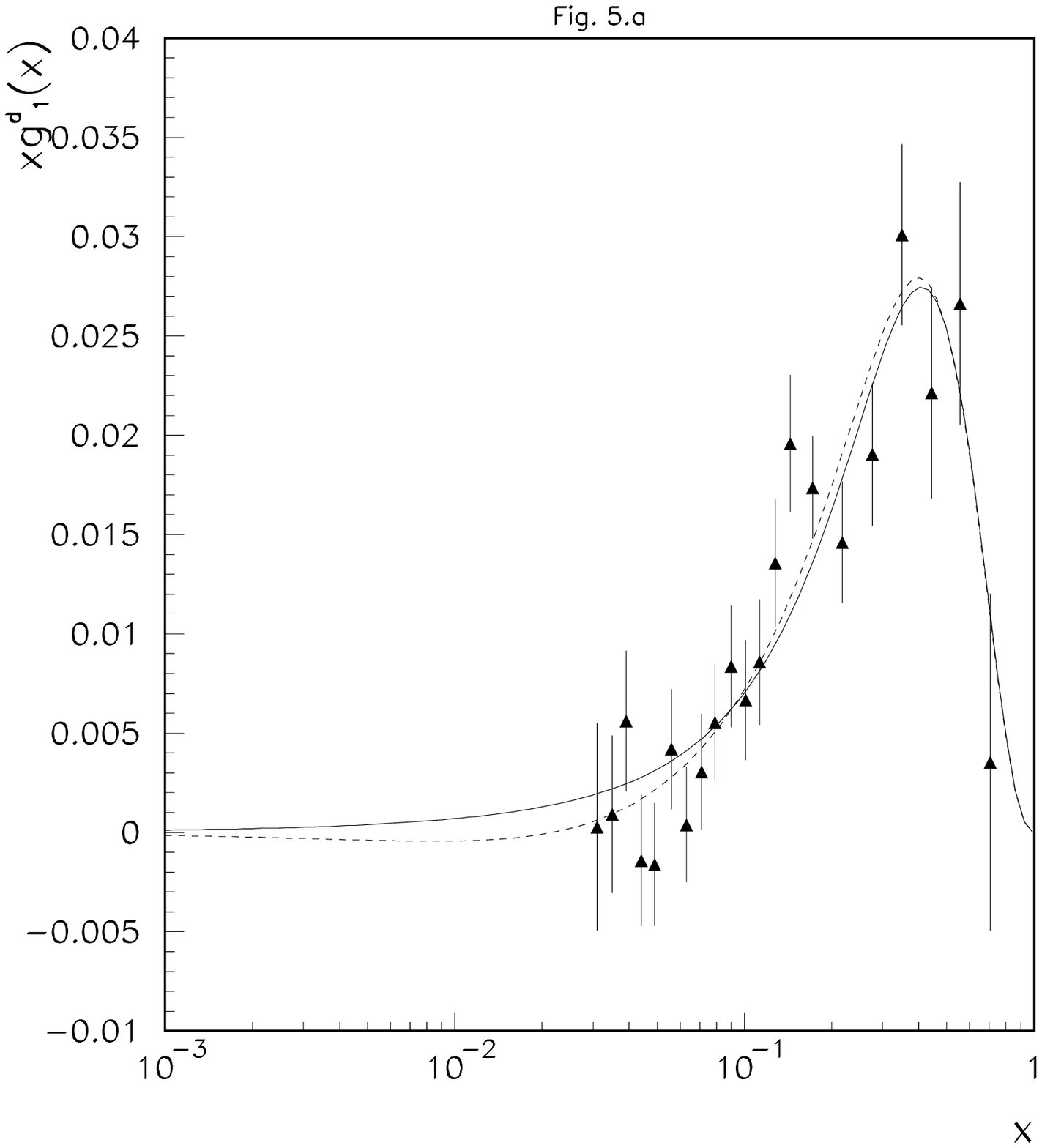,height=9cm}
\epsfig{file=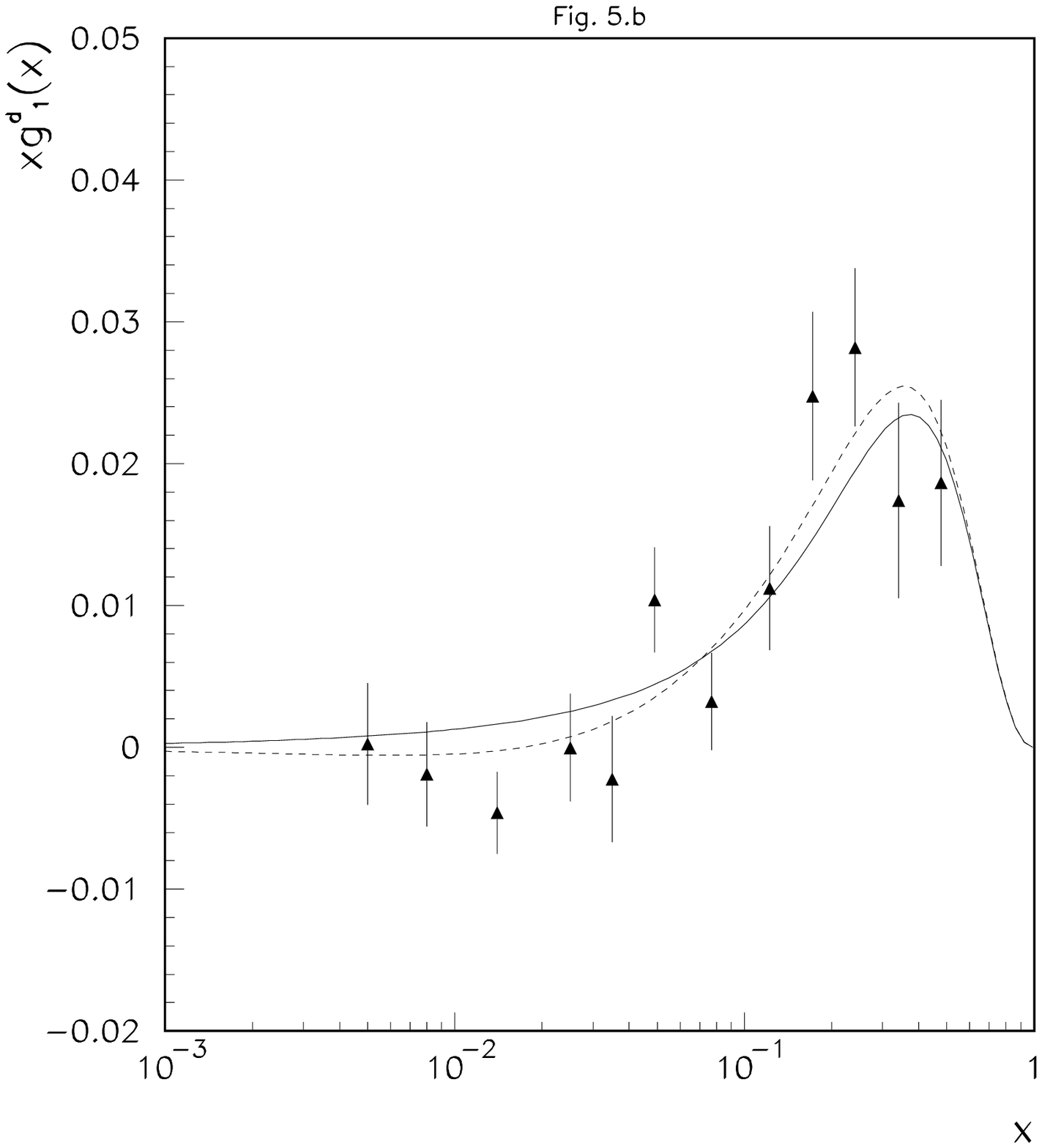,height=9cm}\\
\footnotesize{\begin{itemize}
\item[Figure 5.a] $x\gdx$ at $Q^2=3~(GeV/c)^2$ is plotted and compared with
the data \cite{g1slac}. 
\item[Figure 5.b] The same quantity of Figure 5.a corresponding to
$Q^2=10~(GeV/c)^2$ is plotted versus the experimental data \cite{g1cern}. 
\end{itemize}}

\epsfig{file=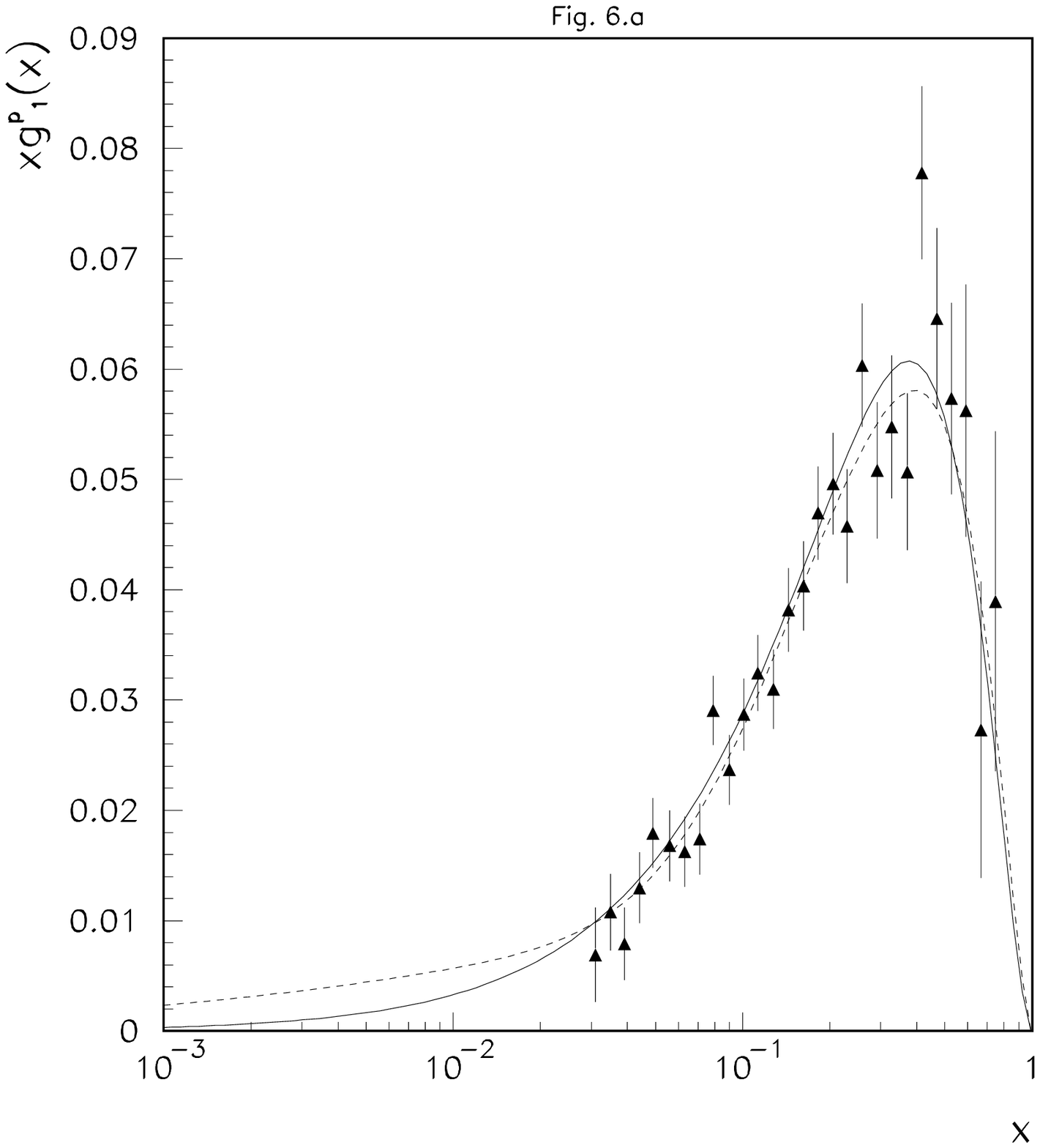,height=9cm}
\epsfig{file=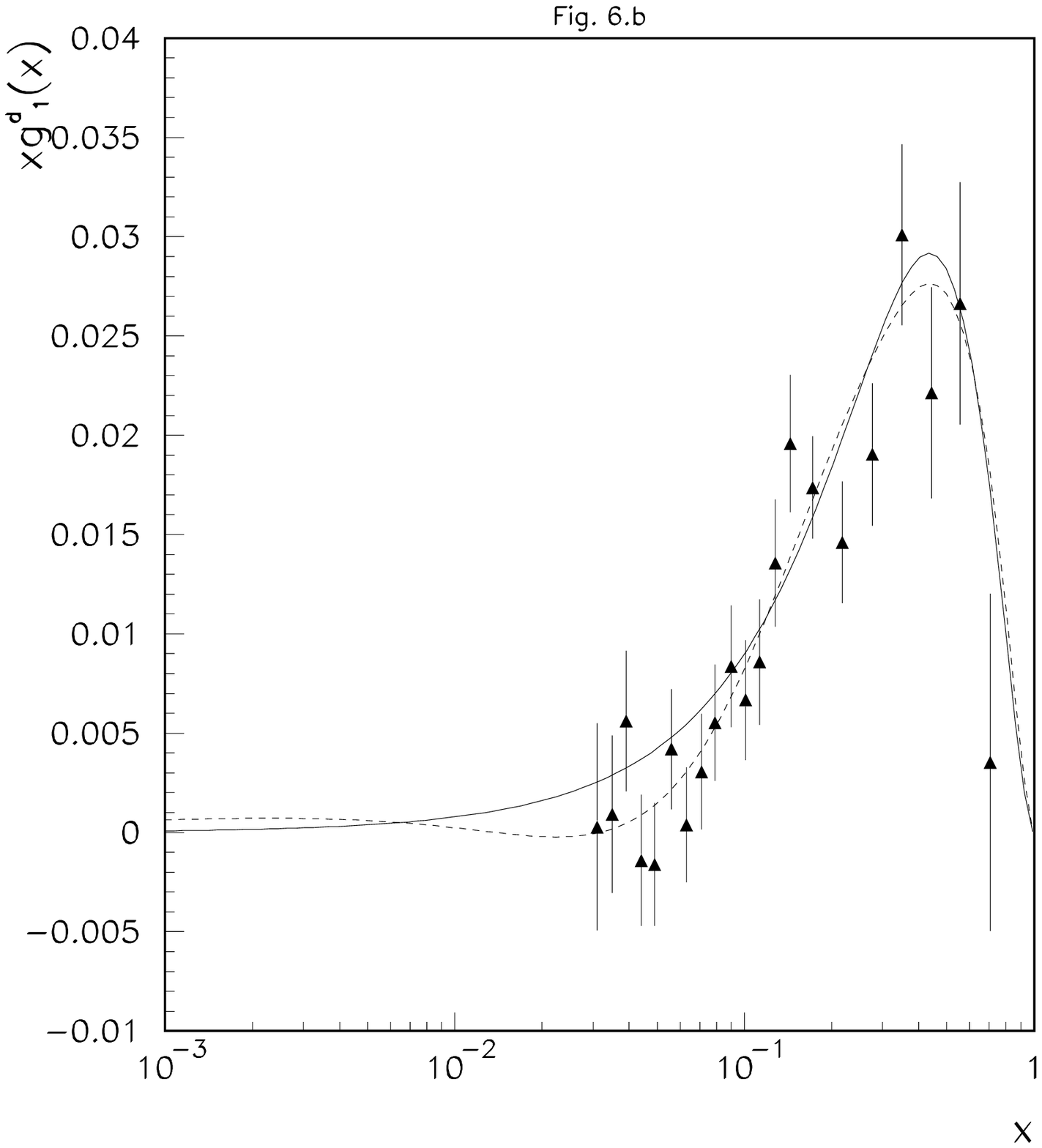,height=9cm}\\
\footnotesize{\begin{itemize}
\item[Fig. 6.a] The best fit for the options {\bf A} (dashed line) and {\bf
B} (solid line) are compared with the SLAC data on proton for $xg^p_1(x)$
at $<\!\!Q^2\!\!> = 3\,(GeV/c)^2$ from ref. \cite{g1slac}. 
\item[Fig. 6.b] Same as Fig. 6.a for the deuteron SLAC data for $xg^d_1(x)$
from ref. \cite{g1slac}. 
\end{itemize}}
\end{figure}
%-----------------------------------------------------------------------
\newpage
\begin{figure}[t]
\centerline{\epsfig{file=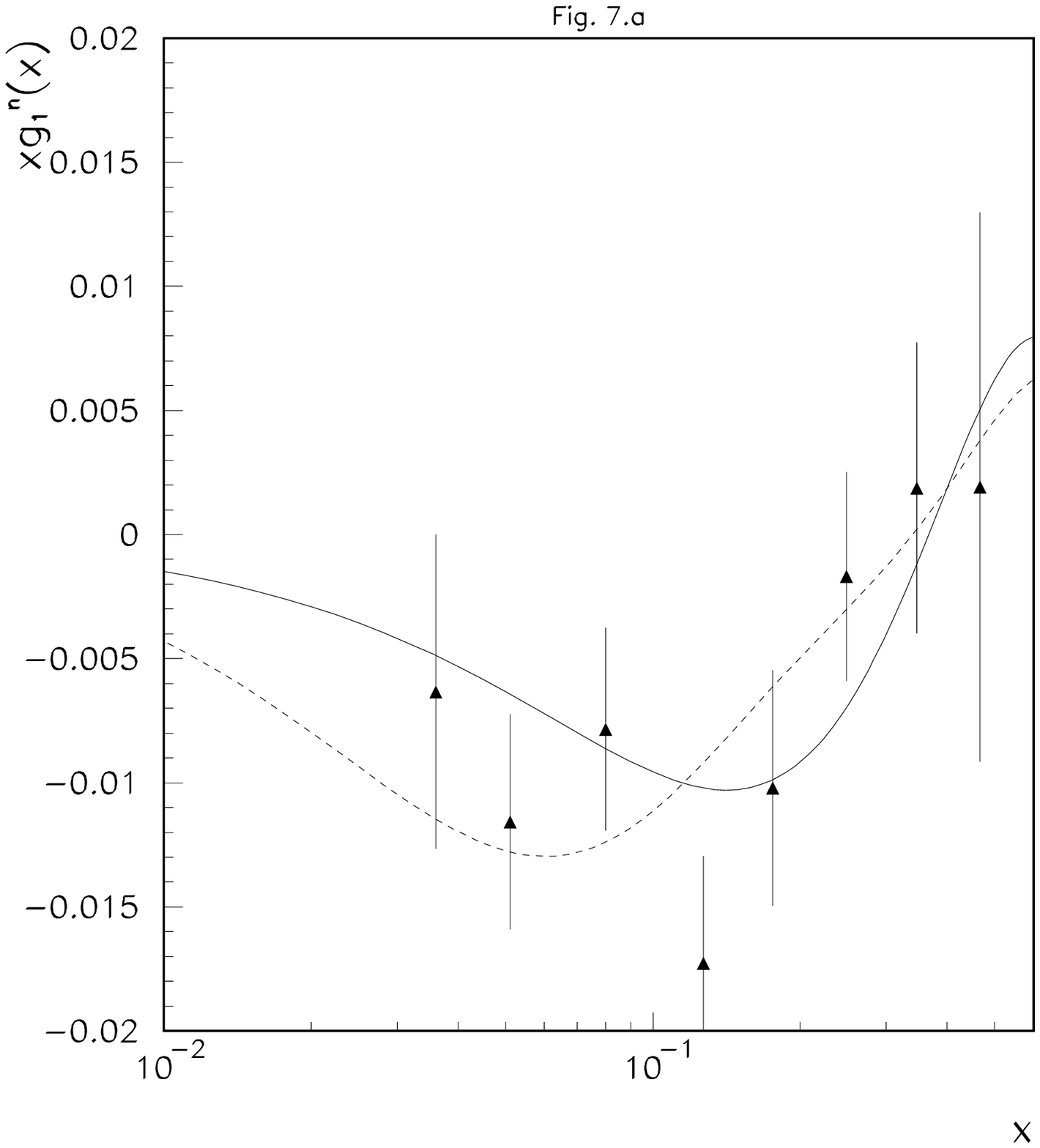,height=9cm}}

\footnotesize{\begin{itemize}
\item[Fig. 7.a] The evolution to $Q^2 = 2\,(GeV/c)^2$ of the results of the
options {\bf A} (dashed line) and {\bf B} (solid line) are compared with
the SLAC-E142 data on neutron for $xg^n_1(x)$ at $<\!\!Q^2\!\!> =
2\,(GeV/c)^2$ from ref. \cite{E142}. 
\end{itemize}}

\epsfig{file=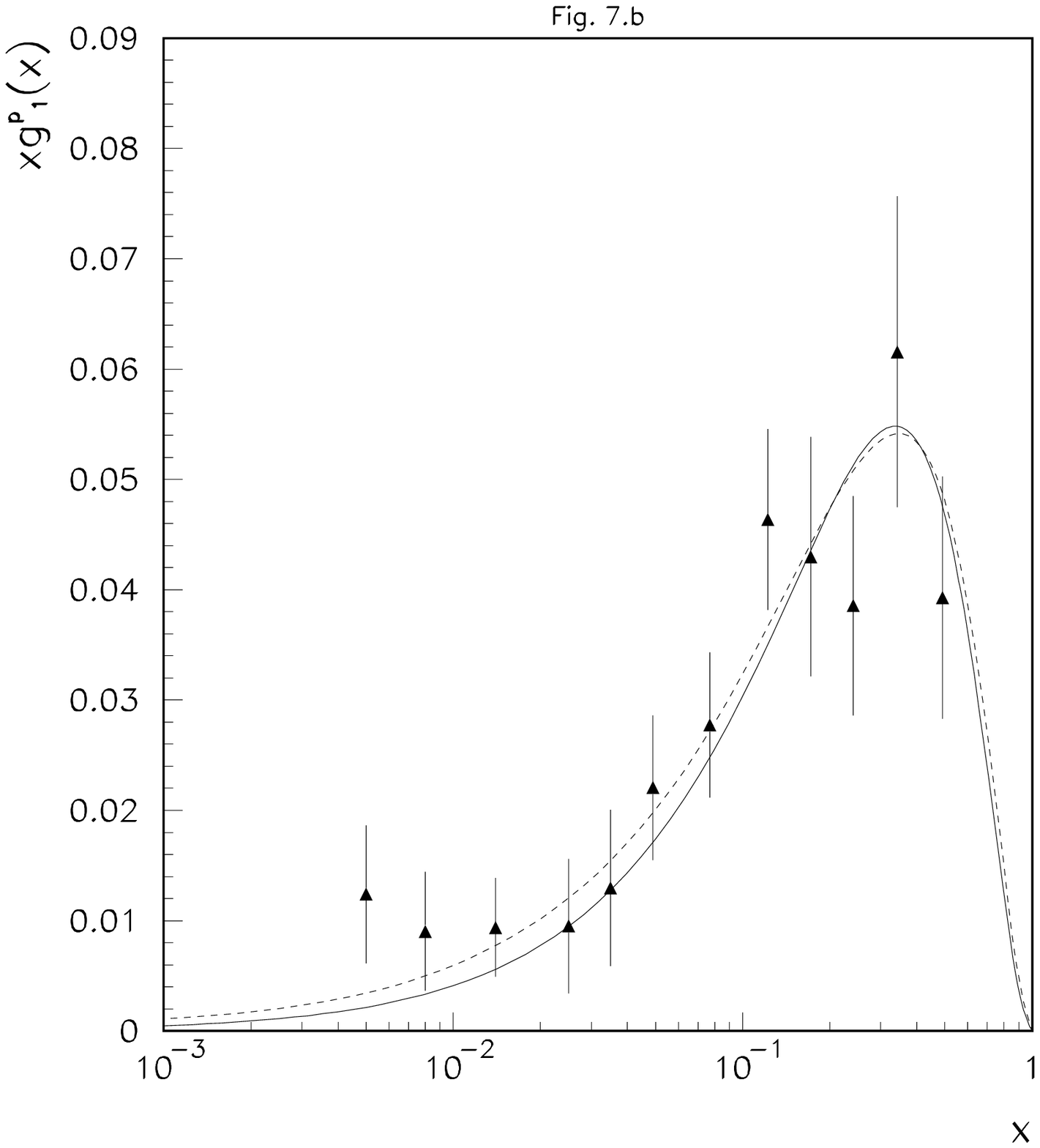,height=9cm}
\epsfig{file=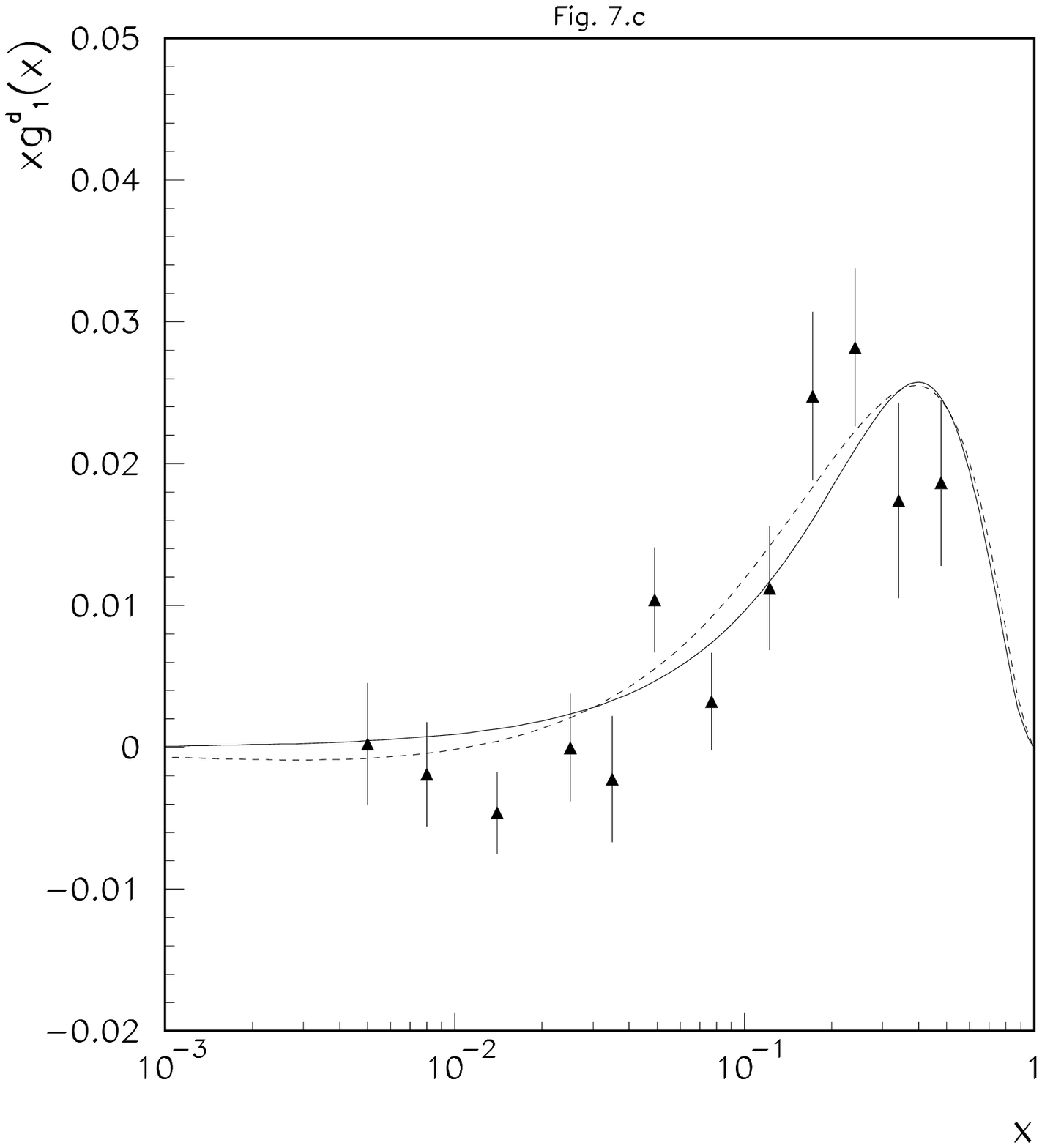,height=9cm}\\
\footnotesize{\begin{itemize}
\item[Fig. 7.b] The data on proton for $xg^p_1(x)$ from SMC at
$<\!\!Q^2\!\!> = 10\,(GeV/c)^2$ from ref. \cite{g1cern} are compared with
the results of the options {\bf A} (dashed line) and {\bf B} (solid line),
evolved to $Q^2 = 10\,(GeV/c)^2$. 
\item[Fig. 7.c] Same as Fig. 4 for the deuteron SMC data for $xg^d_1(x)$
from ref. \cite{g1cern}. 
\end{itemize}}
\label{fig:fig4}
\end{figure}
\newpage
\end{document}